\documentclass[a4paper,11pt]{article}
\pdfoutput=1

\usepackage{jheppub}
\usepackage{amsfonts, amsthm}
\usepackage{amssymb}
\usepackage{amsmath}
\usepackage[english]{babel}
\usepackage[utf8]{inputenc}
\usepackage{slashed}
\usepackage{mathrsfs}
\usepackage{color}
\hypersetup{unicode}
\usepackage{natbib}
\usepackage{lipsum}
\usepackage{tikz}
\usetikzlibrary{arrows.meta}
\usepackage{graphicx}
\usepackage{subcaption}
\usepackage{float}

\usepackage{tabularx}
\usepackage{multirow}
\usepackage[T1]{fontenc}
\usepackage{pifont}

\newcommand{\eq}{\begin{equation}}
\newcommand{\feq}{\end{equation}}
\newcommand{\eqn}{\begin{eqnarray}}
\newcommand{\feqn}{\end{eqnarray}}

\def\Im{{\mathrm{Im}}}
\def\Re{{\mathrm{Re}}}

\def\beq{\begin{align}}
\def\eeq{\end{align}}
\newcommand{\bi}{\begin{itemize}}
\newcommand{\ei}{\end{itemize}}
\newcommand{\ben}{\begin{enumerate}}
\newcommand{\een}{\end{enumerate}}
\providecommand{\bea}{\begin{eqnarray}}
\providecommand{\eea}{\end{eqnarray}}
\newcommand{\be}{\begin{equation}}
\newcommand{\ee}{\end{equation}}

\def\Im{{\mathrm{Im}}}
\def\Re{{\mathrm{Re}}}

\title{Non-SUSY DW's in ISO(7) gauged supergravity}

\author[a]{Nathan Bagshaw} \author[b,c]{Giuseppe Dibitetto}

\affiliation[a]{\'Ecole Normale Sup\`erieure de Lyon, Universit\'e Claude Bernard Lyon I }
\affiliation[b]{Dipartimento di Fisica, Universit\`a di Roma Tor Vergata, via della Ricerca Scientifica 1, 00133 Roma, Italy}
\affiliation[c]{INFN, Sezione di Roma Tor Vergata,via della Ricerca Scientifica 1, 00133 Roma, Italy}

\emailAdd{nathan.bagshaw@ens-lyon.fr, giuseppe.dibitetto@roma2.infn.it}

\abstract{We consider 4D maximal $\mathrm{ISO}(7)$ gauged supergravity, which is known to arise from a consistent truncation of massive IIA supergravity on a six-sphere. Within its SU$(3)$ invariant sector, the theory is known to possess eight AdS vacua, preserving various amounts of residual supersymmetry and bosonic symmetry. By making use of fake supergravity and the Hamilton-Jacobi formalism, we find novel non-supersymmetric domain walls (DW) interpolating between different pairs of AdS extrema. We conclude by discussing some holographically relevant quantities such as the free energy and the anomalous dimensions of the operators triggering the dual RG flows.} 

\begin{document}
\maketitle
\flushbottom

\section{Introduction}
One of the most celebrated developments within string theory is the AdS/CFT correspondence \cite{Maldacena:1997re,Witten:1998qj}, which relates semiclassical gravitational descriptions on AdS to a conformal field theory (CFT) in one less dimension living on the boundary of the AdS spacetime. The holographic dictionary relates bulk field vev's to boundary operators. If one thinks of CFTs as special fixed points of RG flows of non-conformal field theories, it is very natural to extend such a correspondence to gravitational backgrounds in which the AdS isometries are (partially) broken by turning on radial profiles for the bulk fields. A relevant realization of this paradigm is the so-called DW/QFT correspondence \cite{Boonstra:1998mp}, where the RG flow of a non-conformal QFT is mapped into a domain wall (DW) geometry. In this setup, the bulk radial coordinate geometrizes the energy scale of the dual field theory. 

Of special interest in this context is the case where the RG flow interpolates between two different conformal fixed points, one in the IR and one in the UV. The holographic dual geometry is a static DW separating two different AdS spaces, with different radii. Each of the vacua connected by the aforementioned DW may preserve different amounts of supersymmetry, as well as bosonic symmetries. On the other hand, the different mass spectra of the two AdS vacua reflect the anomalous dimension that each operator may develop along the RG flow.

This picture motivates a systematic search for DW solutions that interpolate bewteen different AdS vacua. This is efficiently carried out within those lower dimensional gauged supergravity theories that arise from consistent truncations of string or M theory. However, it is fair to say that the status of such a general notion of holographic correspondence beyond AdS and beyond supersymmetry is still that of a conjecture and there are a few loose ends. First of all, in order for an AdS vacuum to possess a CFT dual, it better be fully stable even at a non-perturbative level. In a metastable vacuum, due to its finite lifetime, Coleman-De Luccia bubbles \cite{Coleman:1980aw} or bubbles of nothing \cite{Witten:1981gj} may spontaneously nucleate, expand and destroy the vacuum. This in turn instantaneously destroys its holographic dual, since a random point in the bulk of AdS is a finite proper distance away from the boundary. 

In cases where a Freund-Rubin type flux is absent in internal space, gravitational instantons such as bubbles of nothing are expected to exist whenever supersymmetry is broken. This is due to the fact that these geometries require anti-periodic boundary conditions for fermions, which are only allowed in non-supersymmetric setup's. In the original construction in \cite{Witten:1981gj}, an internal circle collapses to zero size at the location of the bubble. Generalizations thereof are possible in higher dimensions, where an internal cycle shrinks (see \emph{e.g.} \cite{Ooguri:2017njy}, where a two-cycle collapses). In presence of Freund-Rubin flux, any internal cycle is topologically protected from collapse. However, there are proposals in the literature for different types of decay channels that describe the spontaneous nucleation of a charged bubble hiding a bubble of nothing inside \cite{Horowitz:2007pr,Bomans:2021ara}.

Even when no clear decay channel is detected and even if supersymmetry is unbroken, checking the correspondence explicitly is not an easy task in general, since for some AdS vacua their dual CFT may be non-Lagrangian, like \emph{e.g.} $\mathcal{N}=(2,0)$ CFT$_6$ dual to the $\mathrm{AdS}_7\times S^4$ vacuum of M theory. In presence of non-Lagrangian theories one has to resort to other methods than perturbation theory in order to compute observables and carry out holographic checks. A very valuable tool in this context is the so-called conformal bootstrap (see \emph{e.g.} \cite{Poland:2018epd} for a nice review), which is based on analyzing all consequences of conformal symmetry at the level of the correlators of a CFT, regardless of whether or not it has a Lagrangian description.

By adopting a stringy \emph{top-down} approach, one might say that the AdS constructions that have all rights to be in the landscape are those that can be explicitly obtained by a string compactification within a good perturbative corner, \emph{i.e.} where  higher derivative, quantum loop and instanton corrections are all under control. Even so, there still seem to be spots in the AdS string landscape that do not have a known holographic dual. Emblematic cases of this sort are those AdS vacua featuring scale separation, like the DGKT construction \cite{DeWolfe:2005uu} in massive IIA supergravity, or analogous ones in M theory \cite{Cribiori:2021djm}, or again in three dimensions \cite{Arboleya:2025ocb}. On the CFT side, it appears to be very contrived to have non trivial solutions to the bootstrap equations coming from imposing crossing symmetry, that exhibit a parametrically large separation in the space of conformal dimensions. Indeed, for purely parity preserving 3d CFT's this instance has been ruled out by means of numerical bootstrap techniques. In a fully general case though, this evidence is still lacking and it remains an open issue.

If we focus on those AdS vacua in which supersymmetry is spontaneously broken, the situation does not really get better. 
In the last decade, an entire \emph{String Swampland Program} \cite{Brennan:2017rbf,Palti:2019pca} has been developed with the scope of defining a whole set of  quantum gravity criteria, that could help us tell good stringy EFT's apart from the rest. To start with, these criteria are simply conjectures supported by stringy insights. For our present aim the so-called \emph{AdS Swampland Conjecture} introduced by \cite{Ooguri:2016pdq} turns out be relevant. According to it, every non-supersymmetric AdS vacuum should be ultimately unstable in string theory. The intuition supporting this is the following. If the vacuum at hand is supported by flux, which is generically the case, it will be unstable w.r.t. the spontaneous formation of microscopic membranes that eventually discharge the flux and hence destroy the underlying AdS vacuum. This process is intrinsically non-perturbative and does not occur in supersymmetric vacua, since the spectrum of excitations \emph{only} contains BPS membranes, for which the tension equals the charge. In \cite{Danielsson:2016mtx} it was argued that such instabilities may occur at a perturbative level when the effective supergravity description is coupled to the open string sector. Later in \cite{Danielsson:2017max}, this proposal was explicitly checked in a few cases of interest.

A straightforward consequence of the AdS swampland conjecture would simply be that non-supersymmetric holography does not exist. In this paper we study (non-)supersymmetric $\mathrm{AdS}_4\times S^6$ vacua of massive IIA supergravity. Such solutions are captured by a 4D gauged supergravity theory with gauge group $\mathrm{ISO}(7)$ \cite{Guarino:2015qaa}, and in particular its $\mathrm{SU}(3)$ invariant subsector. Remarkably, a few non-supersymmetric AdS vacua in this setup have already passed all perturbative checks, including the absence of tachyons within the full KK spectrum \cite{Guarino:2020flh}, and the absence of brane jet instabilities of different kinds \cite{Guarino:2020jwv}. Possible other non-perturbative instability channels remain to be checked. The scope of our present work is to study whether a non-supersymmetric DW/QFT correspondence might instead yield meaningful results here. This should be seen as a stress test for the AdS swampland conjecture. 

In this paper we study the problem within a 4D $\mathcal{N}=1$ supergravity model capturing both supersymmetric as well as  non-supersymmetric AdS extrema. Following the original idea in \cite{Freedman:2003ax}, we make use of the Hamilton-Jacobi (HJ) formalism as our main technical tool. The appropriate HJ generating functions play the role of \emph{fake superpotentials} and define first order flow equations for all DW solutions, interpolating between different pairs of critical points. Each of these profiles, if one may trust in the holographic dictionary, identifies an RG flow within a 3d field theory. The present work may be regarded as a generalization of the analysis carried out in \cite{Dibitetto:2021kcz} within the G$_2$ invariant sector of the theory, to its SU$(3)$ invariant one.

The paper is organized as follows. In section \ref{Sec:SU3_truncation} we introduce $\mathrm{ISO}(7)$ gauged maximal supergravity, its truncation to the $\mathrm{SU}(3)$ invariant sector and its vacuum structure. In section \ref{Sec:Flat_DWs} we discuss first order flows by using the HJ formalism in the context of fake supergravity. In particular, we review how the fake superpotential generating the HJ flow may be determined by a perturbative expansion around any AdS critical point of the scalar potential. In section \ref{Sec:DW_Plots} we present our interpolating static DW solutions and illustrate them with a series of plots. In section \ref{Sec:Holography} we present a few holographic considerations and make some concluding remarks.
Finally, we present in appendix \ref{AppA} some technical details concerning the massive type IIA supergravity origin of our 4D effective model.

\section{The $\mathrm{SU}(3)$ invariant sector of $\mathrm{ISO}(7)$ gauged maximal supergravity}
\label{Sec:SU3_truncation}
$\mathrm{ISO}(7)$ gauged $\mathcal{N}=8$ supergravity is a consistent gauged maximal supergravity in $D=4$ belonging to the $\mathrm{CSO}(p,q,8-p-q)$ class, \emph{i.e.} all gauge groups obtained from $\mathrm{SO}(8)$ \cite{Cremmer:1979up} by performing a chain of analytic continuations and In\"on\"u-Wigner contractions \cite{deWit:2007kvg}. In these theories the gauge group is embedded in the $\mathrm{E}_{7(7)}$ global symmetry of the ungauged theory through its maximal $\mathrm{SL}(8,\mathbb{R})$ subgroup. The corresponding embedding tensor of the theory lies within the $\textbf{36}\oplus\overline{\textbf{36}}$ of $\mathrm{SL}(8,\mathbb{R})$:
\be
\begin{array}{cccc}
\mathrm{E}_{7(7)} & \supset & \mathrm{SL}(8,\mathbb{R}) & , \\
\textbf{912} & \rightarrow & \underbrace{\textbf{36}}_{Q^{(MN)}}\oplus\underbrace{\overline{\textbf{36}}}_{P_{(MN)}}\oplus\,\textbf{420}\oplus\overline{\textbf{420}} & ,
\end{array}
\ee
All consistent $\mathrm{CSO}(p,q,8-p-q)$ gaugings are then parametrized by the two symmetric $8\times 8$ matrices $Q$ \& $P$ that satisfy the following quadratic constraint
\be\label{QC}
\mathrm{Tr}(QP) \ = \ Q^{MN}\,P_{MN} \ \overset{!}{=} \, 0 \ .
\ee
The purely electric $\mathrm{ISO}(7)$ theory arising from the reduction of \emph{massless} IIA supergravity on a six-sphere \cite{Hull:1988jw}, corresponds to the choice 
\be
Q \ = \ \left(\begin{array}{c|c}0& \\ \hline & q \,\mathbb{I}_7\end{array}\right) \qquad \textrm{and} \qquad P \ = \ \mathbb{O}_8 \ .
\ee
However, more general dyonic $\mathrm{ISO}(7)$ gaugings may be obtained by turning on an extra parameter which takes a symplectic deformation \cite{DallAgata:2014tph} into account. This yields 
\be
Q \ = \ \left(\begin{array}{c|c}0& \\ \hline & q \,\mathbb{I}_7\end{array}\right) \qquad \textrm{and} \qquad P \ = \ \left(\begin{array}{c|c}p& \\ \hline & \mathbb{O}_7\end{array}\right) \ ,
\ee
which still solves the quadratic constraint \eqref{QC} just as the electric $\mathrm{ISO}(7)$ theory did. In type IIA language, the new $p$ parameter controlling the symplectic deformation turns out to be related to the Romans' mass. Hence, the theory originates from a consistent truncation of \emph{massive} IIA supergravity on a six-sphere \cite{Guarino:2015vca}. Some details concerning the truncation to its $\mathrm{SU}(3)$ invariant sector are collected in appendix.
As we will see later on, contrary to its massless counterpart, this theory contains non-trivial AdS$_4$ vacua.

The $32$ real supercharges of the $\mathcal{N}=8$ theory transform in the fundamental representation of the $\mathrm{SU}(8)$ R-symmetry group. If one restricts to the $\mathrm{SU}(3)$ invariant sector, these decompose as
\be
\begin{array}{ccccccccc}
\mathrm{SU}(8) & \supset & \mathrm{SO}(8) & \supset & \mathrm{SO}(7)_+ & \supset & \mathrm{G}_2 & \supset & \mathrm{SU}(3) \\[2mm]
\textbf{8} & \rightarrow & \textbf{8}_v & \rightarrow & \textbf{8} & \rightarrow & \textbf{1} \oplus \textbf{7} & \rightarrow &  \textbf{1} \oplus \textbf{1} \oplus \textbf{3} \oplus \overline{\textbf{3}}
\end{array}
\ee
As a consequence, the truncated theory enjoys $\mathcal{N}=2$ supersymmetry. The 70 propagating scalar degrees of freedom of maximal supergravity decompose as
\be
\hspace{-3mm}
\begin{array}{ccccccccc}
\textbf{70} & \rightarrow & \textbf{35}_v  \oplus \textbf{35}_s & \rightarrow &  \textbf{1} \oplus \textbf{7} \oplus \textbf{27} \oplus \textbf{35} & \rightarrow & 2 \times\left(\textbf{1} \oplus \textbf{7} \oplus \textbf{27} \right) & \rightarrow & 6\times  \textbf{1} \oplus \textrm{non-singlets} \ ,
\end{array}
\ee
where the $\mathrm{SU}(3)$ singlet scalars are organized into one complex field associated with one vector multiplet and four real fields associated with one hypermultiplet. The scalar coset geometry is then factorized into a special K\"ahler (SK) part and a quaternionic  K\"ahler (QK) part through
\be
\label{scalarmanifold}
\mathcal{M}_{\textrm{scalar}} \ = \ \mathcal{M}_{\textrm{SK}}  \ \times \ \mathcal{M}_{\textrm{QK}} \ = \ 
\frac{\mathrm{SL}(2,\mathbb{R})}{\mathrm{SO}(2)} \, \times \, \frac{\mathrm{SU}(2,1)}{\mathrm{U}(2)}\ .
\ee
%r

When further specifiying the embedding tensor for the $\mathrm{ISO}(7)$ gauging, one finds that the corresponding scalar potential only depends on two gauge-invariant real scalars within the universal hypermultiplet. Through this operation, one effectively breaks the quaternionic geometry into another $\frac{\mathrm{SL}(2,\mathbb{R})}{\mathrm{SO}(2)}$ factor \cite{Guarino:2015tja}. This gauged-fixed version of the theory only turns out to admit an $\mathcal{N}=1$ superpotential formulation. For many purposes it may be useful to view our $\mathrm{SU}(3)$ invariant theory as a special sector of a truncation w.r.t. a discrete $\mathbb{Z}_2^3$ symmetry.

This $\mathbb{Z}_2^3$ truncation \cite{Guarino:2020flh,Bobev:2020qev} is fully characterized by assigning a consistent set of parities to the different components of a vector transforming in the fundamental representation of $\mathrm{SL}(8,\mathbb{R})$
\be
\begin{array}{lclclc}
\mathbb{Z}_2^{(1)} & : & \left(x_1,x_2,x_3,x_4,x_5,x_6,x_7,x_8\right) & \longrightarrow &  \left(x_1,x_2,x_3,-x_4,-x_5,-x_6,-x_7,x_8\right) & ,\\[2mm]
\mathbb{Z}_2^{(2)} & : & \left(x_1,x_2,x_3,x_4,x_5,x_6,x_7,x_8\right) & \longrightarrow &  \left(x_1,-x_2,-x_3,x_4,x_5,-x_6,-x_7,x_8\right) & ,\\[2mm]
\mathbb{Z}_2^{(3)} & : & \left(x_1,x_2,x_3,x_4,x_5,x_6,x_7,x_8\right) & \longrightarrow &  \left(x_1,-x_2,x_3,-x_4,x_5,-x_6,x_7,-x_8\right) & .
\end{array}
\ee
The theory obtained by this procedure enjoys $\mathcal{N}=1$ supersymmetry and contains seven chiral multiplets, the corresponding scalar geometry being
\be
\left\{\Phi_I\right\}_{I=1,\dots,7} \ \in \ \left(\frac{\mathrm{SL}(2,\mathbb{R})}{\mathrm{SO}(2)}\right)^7 \ .
\notag
\ee
The general K\"ahler potential determining the scalar kinetic coupling reads
\be
K(\Phi,\bar{\Phi}) \, = \, -\sum\limits_{I=1}^7 \log\left(-i(\Phi_I-\bar{\Phi}_I)\right) \ .
\ee
 The $\mathrm{ISO}(7)$ embedding tensor deformations generate the following superpotential
\be
W \, = \, 2g\left(c\,+\,\Phi_1\Phi_2\Phi_3+ \Phi_1\left(\Phi_4\Phi_5+\Phi_6\Phi_7\right)+\Phi_2\left(\Phi_4\Phi_6+\Phi_5\Phi_7\right)+\Phi_3\left(\Phi_4\Phi_7+\Phi_5\Phi_6\right)\right) \ .
\ee

The corresponding action is given by 
\be
\label{action_Z23}
\mathcal{S}_{\mathbb{Z}_2^3}[g_{\mu\nu},\Phi,\bar{\Phi}] \, = \,\int{d^4x\sqrt{-g_4}\,\left(\frac{\mathcal{R}_4}{2\kappa_4^2}-K_{I\bar{J}}\,(\partial \Phi^I) (\partial \bar{\Phi}^{\bar{J}}) \, - \, V(\Phi,\bar{\Phi})\right)} \ ,
\ee
where $\kappa_4$ denotes the 4D gravitational coupling, $K_{I\bar{J}}\,\equiv\,\partial_I\partial_{\bar{J}}K$, and the scalar potential $V$ is determined by the superpotential through\footnote{From now on we set $\kappa_4=1$.}
\be
V \ = \ e^{K} \, \left(-3|W|^2\,+\,K^{I\bar{J}}D_IWD_{\bar{J}}\bar{W}\right) \ ,
\label{Potential}
\ee
where $D$ denotes the K\"ahler covariant derivative, \emph{e.g.} $D_{I}W \, \equiv \, \partial_IW +\partial_I K\, W$, and $K^{I\bar{J}}$ indicate the components of the inverse K\"ahler metric. Models with enhanced residual symmetry may be understood as special sectors of the $\mathbb{Z}_2^3$ truncation. The corresponding explicit identifications for some relevant cases are collected in table \ref{Table:Truncations}.
\begin{table}[http!]
\renewcommand{\arraystretch}{1}
\begin{center}
\scalebox{1}[1]{
\begin{tabular}{|c| c | c|}
\hline
$G_{\textrm{res.}}$ & $\left\{\Phi_I\right\}_{I=1,\dots,7}$ parametrization & \# (real) d.o.f.'s  \\ \hline \hline
$\mathrm{O}(3)$ & $(T,T,T,S,U,U,U)$ &  $6$ \\ \hline
$\mathrm{SO}(4)$ & $(T,T,T,S,T,T,T)$ &  $4$ \\ \hline
$\mathrm{SU}(3)$ & $(T,T,T,S,S,S,S)$ & $4$  \\ \hline
$\mathrm{SU}(3)\times \mathrm{U}(1)$ & $(T,T,T,S,S,S,S)$, with $\mathrm{Re}(S)=0$ &  $3$ \\ \hline
$\mathrm{SO}(6)_+$ & $(T,T,T,S,S,S,S)$, with $\mathrm{Re}(S)=\mathrm{Re}(T)=0$  &  $2$ \\ \hline
${\mathrm{G}_2}$ & $(S,S,S,S,S,S,S)$ &  $2$  \\ \hline
 $\mathrm{SO}(7)_+$ & $(S,S,S,S,S,S,S)$, with $\mathrm{Re}(S)=0$ &  $1$\\
\hline 
\end{tabular}
}
\end{center}
\caption{\it Summary of group theoretical truncations of ISO(7) gauged supergravity in the language of minimal SUSY STU models.} \label{Table:Truncations}
\end{table}

If we now focus on the $\mathrm{SU}(3)$ invariant sector (see ref \cite{Varela_2016} where this sector was exhaustively explored), we may describe its dynamics in terms of the following superpotential
\be
W_{\mathrm{SU}(3)} \,=\, 2g \left(c+S^3+6S^2T\right)\ .
\label{W_SU(3)}
\ee
In the remainder of our analysis we shall set $g\,=\,c\,=\,1$ as a choice of units.
We fix the following specific scalar parametrization in terms of real fields
\be
S\,=\, s \, + \, i \sigma \ , \quad \textrm{and} \qquad T\,=\, t \, + \, i \tau \ .
\ee
The explicit expression of the scalar potential then reads
\begin{multline}
\label{pot}
V = \frac{1}{8 \tau^3 \sigma^4} \big( 
\tau^6 + t^6 + 3 \tau^4 (t^2 - 4 \sigma^2) + 12 t^4 s^2 + 
36 t^2 s^2 (\sigma^2 + s^2)  \\
+ 3 \tau^2 \left( t^4 - 8 \sigma^4 - 4 \sigma^2 s^2 + 4 s^4 - 4 t^2 (\sigma^2 - s^2) \right) + 
2 t^3 + 12 t s^2 + 1 
\big)~.
\end{multline}
This potential admits eight different critical points, some of which are supersymmetric, while the remaining ones have sponteneously broken SUSY. With our choice of embedding tensor normalization, the aforementioned critical points are illustrated in table \ref{Table:Critical_Points}, while the corresponding mass spectra all shown in table \ref{Table:Mass_spectra}.
\begin{table}[http!]
\centering
\begin{tikzpicture}

\renewcommand{\arraystretch}{1.4}
  % Met le tableau dans un node
  \node (tab) at (0,0) {
    \begin{tabular}{|c||c|c|c|c|}
\hline
Id & $G_{\text{res}}$ & SUSY & $(s,\sigma,t,\tau)$ & $V_0$ \\
\hline \hline
1 & $\mathrm{SO}(7)_+$ & $\mathcal{N}=0$ & $\left(0,\tfrac{1}{5^{1/6}},0,\tfrac{1}{5^{1/6}}\right)$ & $-\tfrac{3\cdot 5^{7/6}}{4}$ \\
\hline
2 & $\mathrm{G}_2$ & $\mathcal{N}=1$ & $2^{-7/3}\left(1,\sqrt{15},1,\sqrt{15}\right)$ & $-\frac{2^{22/3}\sqrt{3}}{5^{5/2}}$ \\
\hline
3 & $\mathrm{SU}(3)\times \mathrm{U}(1)$ & $\mathcal{N}=2$ & $\left(0,\tfrac{1}{\sqrt{2}},\tfrac{1}{2},\tfrac{\sqrt{3}}{2}\right)$ & $-3^{3/2}$ \\
\hline
4 & $\mathrm{SO}(6)_+$ & $\mathcal{N}=0$ & $\left(0,\tfrac{1}{2^{5/6}},0,\tfrac{2}{2^{5/6}}\right)$ & $-3\cdot 2^{5/6}$ \\
\hline
5 & $\mathrm{G}_2$ & $\mathcal{N}=0$ & $2^{-4/3}\left(-1,\sqrt{3},-1,\sqrt{3}\right)$ & $-\frac{2^{10/3}}{\sqrt{3}}$ \\
\hline
6 & $\mathrm{SU}(3)$ & $\mathcal{N}=0$ & $\left(0.4915,0.6618,-0.2698,0.7329\right)$ & $-5.8534$ \\
\hline
7 & $\mathrm{SU}(3)$ & $\mathcal{N}=0$ & $\left(0.3348,0.6012,-0.4546,0.8376\right)$ & $-5.8642$ \\
\hline
8 & $\mathrm{SU}(3)$ & $\mathcal{N}=1$ & $\tfrac{1}{4}\left(\sqrt{3},\sqrt{5},-1,\sqrt{15}\right)$ & $-\tfrac{3^{3/2}\cdot 2^6}{5^{5/2}}$ \\
\hline
\end{tabular}
  };

  % Ajoute une flèche verticale à droite
  \draw[->, thick] ([xshift=0.6cm, yshift=-3cm]tab.east) -- ++(0,5.8) node[right] {~$V_0$};

\end{tikzpicture}
\caption{\it The eight different AdS critical points of $\mathrm{ISO}(7)$ gauged supergravity preserving at least an $\mathrm{SU}(3)$ residual symmetry, arranged by decreasing values of the on-shell potential.  }
\label{Table:Critical_Points}
\end{table}

\newcommand{\cmark}{\ding{51}} 
\newcommand{\xmark}{\ding{55}}

\begin{table}[h!]
\centering
\hspace*{-1.5cm}
\renewcommand{\arraystretch}{1.2}
\begin{tabular}{|c||c|c|c|c|}
\hline
\multirow{2}{*}{Id} &\multirow{2}{*}{ masses $m^2L^2$ }& \multirow{2}{*}{modes $\Delta$} & BF & KK \\
 &  & & stability& stability \\
\hline \hline
1 & $6,~ -\tfrac{12}{5}, ~-\tfrac{6}{5} (\times 2)$ & $\tfrac{1}{2}(3 \pm \sqrt{33}), \tfrac{1}{10}(15 \pm \sqrt{105})~(\times 2)$ & \xmark & \xmark \\
\hline
\multirow{2}{*}{2} & \multirow{2}{*}{$4\pm\sqrt{6},~ \tfrac{1}{6}(-11\pm\sqrt{6})$} & $1 \pm \sqrt{6},~2 \pm \sqrt{6},~$ & \multirow{2}{*}{\cmark} & \multirow{2}{*}{\cmark}\\
 & & $\tfrac{1}{6}(6 \pm \sqrt{6}),~ \tfrac{1}{6}(12 \pm \sqrt{6})$ && \\
\hline
 3& $3\pm \sqrt{17},~ 2(\times 2)$ & $\tfrac{1}{2}(1\pm \sqrt{17}),~ \tfrac{1}{2}(3\pm \sqrt{17})(\times 2),~\tfrac{1}{2}(5\pm \sqrt{17})$ & \cmark & \cmark \\
\hline
4 & $6(\times 2),~ 0,~ -\tfrac{3}{4}$ & $\textcolor{red}{0},~3,~ \textcolor{red}{\tfrac{1}{2} (3} \raisebox{0.06cm}{\ooalign{$+$\cr\raisebox{-0.15cm}{\textcolor{red}{$-$}}\cr}} \textcolor{red}{\sqrt{6})},~ \textcolor{red}{\tfrac{1}{2}(3} \raisebox{0.06cm}{\ooalign{$+$\cr\raisebox{-0.15cm}{\textcolor{red}{$-$}}\cr}}\textcolor{red}{\sqrt{33})~(\times 2)} $ & \cmark & \xmark  \\
\hline
5 & $6(\times 2),~ -1(\times 2)$ & $\tfrac{1}{2}(3 \pm \sqrt{5})~(\times 2),~ \tfrac{1}{2}(3 \pm \sqrt{33})~(\times 2)$ & \cmark & \cmark \\
\hline
\multirow{2}{*}{6} & \multirow{2}{*}{$6.2296, 5.9049, -1.2644, 1.1299$} & $\textcolor{blue}{-1.41197, -1.35568, -0.338452, 0.507214},$ & \multirow{2}{*}{\cmark}& \multirow{2}{*}{\cmark } \\
 & & $2.49279, 3.33845, 4.35568, 4.41197$ && \\
\hline
\multirow{2}{*}{7} & \multirow{2}{*}{$6.2144, 5.9251, -1.2844, 1.1448$} & $\textcolor{orange!92!black}{-1.40937, -1.35921, -0.3425, 0.517326},$ & \multirow{2}{*}{\cmark}& \multirow{2}{*}{\cmark} \\
 & & $2.48267, 3.3425, 4.35921, 4.40937$ && \\
\hline
8 & $4\pm\sqrt{6} ~(\times 2)$ & $\textcolor{green!60!black}{1} \raisebox{0.06cm}{\ooalign{$+$\cr\raisebox{-0.15cm}{\textcolor{green!60!black}{$-$}}\cr}}\textcolor{green!60!black}{\sqrt{6}~(\times 2)} ,~\textcolor{green!60!black}{2} \raisebox{0.06cm}{\ooalign{$+$\cr\raisebox{-0.15cm}{\textcolor{green!60!black}{$-$}}\cr}}\textcolor{green!60!black}{\sqrt{6}~(\times 2)}$ & \cmark & \cmark  \\
\hline
\end{tabular}

\caption{\it For the different AdS vacua in table \protect\ref{Table:Critical_Points}, we display the squared masses of the $\mathrm{SU}(3)$ singlet modes, normalized to the AdS radius $L$. In these units the BF bound \protect\cite{Breitenlohner:1982jf} lies at $m^2_{\textrm{BF}}L^2=-\frac{9}{4}$. The results concerning the stability of the whole KK tower were found in \protect\cite{Guarino:2020flh}.}
\label{Table:Mass_spectra}
\end{table}

The mass spectra displayed in table \ref{Table:Mass_spectra} are obtained as the eigenvalues of the mass matrix $(m^2L^2)^{i}_{~k}\equiv-\tfrac{3K^{ij}V_{jk}^{(2)}}{V_0}$. We remind the reader that in AdS vacua, small negative values for the squared masses are acceptable as long as they respect the BF bound (here $m^2L^2\geq -\tfrac{9}{4}$). Within our truncation only the SO$(7)_+$ vacuum exhibits a tachyon, making it unstable. However, we shall later see that this does not prevent one from finding DWs connecting this point to others, despite its instability. In order to better understand this, let us probe the stability of a vacuum by checking the linearized flow equation \eqref{HJ_flow_eqns} (see later), with $\phi^i(z)=\phi^i_0+\delta \phi^i(z)$
\begin{equation}
(\delta \phi^j)'=-2K^{ij}|_{\phi_0}\left.\frac{\partial^2 f}{\partial\phi^k\partial\phi^i}\right|_{\phi_0}\delta \phi^k ~. 
\label{lineareq}
\end{equation}
This is a first order differential equation, hence the perturbation can be written in the form \begin{equation}
\delta \phi^j = \zeta^j_{~k}e^{-\frac{z}{L}\Delta_k} ~ ,
\end{equation}
where $\Delta_k $ are the eigenvalues of the matrix $T\equiv 2Lf^{(2)}K^{-1}$. On the other hand, $f^{(2)}$ is related to the Hessian matrix for the scalar potential $V^{(2)}$. Hence we have
\begin{equation}
\begin{split}
&f^{(2)}=\frac{3}{4L}K+\frac{1}{2}K^{1/2}\sqrt{\frac{9}{4L^2}+K^{1/2}m^2K^{-1/2}}K^{1/2} \\
\Rightarrow ~&T=\frac{3}{2}+K^{1/2}\sqrt{\frac{9}{4}+K^{1/2}m^2L^2K^{-1/2}}K^{-1/2}~.
\end{split}
\end{equation}
$T$ has the same eigenvalues as $\frac{3}{2}+\sqrt{\frac{9}{4}+m^2L^2}$ (because $K$ is diagonal) which is the exact solution to $m^2L^2= \Delta (\Delta -3)$ meaning that the modes $\Delta_k$ and the mass eigenvalues $m_k^2L^2$ verify the relation \cite{Zaffaroni:2000vh}
\begin{equation}
m^2_kL^2= \Delta_k (\Delta_k -3)\ .
\label{delta}
\end{equation}
This relation is the core of the AdS/CFT correspondence and it identifies  the scaling dimension $\Delta$ of the dual boundary operator to the corresponding bulk scalar field. For $\Delta$ to be real, one must impose $m^2L^2\geq -\tfrac{9}{4}$. 
%In order to obtain the modes that are listed next to the masses in Table.\ref{tableaumass}, we have simply solved this $2^{\mathrm{nd}}$ order polynomial. Notice that the polynomial gives us twice as much solutions than the number of fields, which corresponds to the $2^N$ branches discussed in \ref{perturbation}. A fake superpotential corresponds to some specific choice of mode for each scalar. In particular, DWs are solutions that excite the most negatives modes.

\section{Flat first order flows: separable HJ treatment}
\label{Sec:Flat_DWs}
Static DW solutions within the $\mathrm{SU}(3)$ invariant sector of $\mathrm{ISO}(7)$ gauged supergravity may be studied by exploiting the first order formulation through the HJ formalism. These flows are solutions of the form
\be
\label{FlatDW_Ansatz}
ds_4^2 \, = \, e^{2A}ds^2_{\textrm{Mkw}_3} \, + \,dz^2 \ ,
\ee
where $A$, as well as the scalar fields $(s,\sigma,t,\tau)$, are functions of the radial coordinate $z$.

When plugging the \emph{Ansatz} \eqref{FlatDW_Ansatz} into the action \eqref{action_Z23}, one obtains the following 1D effective Lagrangian\footnote{This form of the effective 1D Lagrangian is actually obtained by performing an integration by parts in order to get rid of the term with the second derivative $A''$.}
\be
\label{Leff_FlatDW}
L_{\textrm{eff}} \,=\, e^{3A}\left(3(A')^2-\frac{1}{\sigma ^2}((s')^2+(\sigma ')^2)-\frac{3}{4\tau ^2}((t')^2+(\tau ')^2)-V(s,\sigma ,t,\tau)\right)\ ,
\ee
where $\prime$ denotes differentiation w.r.t. $z$, and $V$ is the scalar potential introduced in \eqref{Potential}.
Now we introduce the conjugate momenta for our dynamical fields $(A,\,s,\,\sigma,\,t,\,\tau)$
\be\left\{
\begin{array}{lclc}
\pi_{A} & = & \dfrac{\partial L_{\textrm{eff}}}{\partial A'} \, = \,  6\,e^{3A}A'& , \\[3mm]
\pi_{s} & = & \dfrac{\partial L_{\textrm{eff}}}{\partial s'} \, = \, -\dfrac{2}{\sigma^2}e^{3A}s' & , \\[3mm]
\pi_{\sigma} & = & \dfrac{\partial L_{\textrm{eff}}}{\partial \sigma '} \, = \, -\dfrac{2}{\sigma^2}e^{3A}\sigma' & , \\[3mm]
\pi_{t} & = & \dfrac{\partial L_{\textrm{eff}}}{\partial t'} \, = \, -\dfrac{3}{2\tau^2}e^{3A}t' & , \\[3mm]
\pi_{\tau} & = & \dfrac{\partial L_{\textrm{eff}}}{\partial \tau'} \, = \,  -\dfrac{3}{2\tau^2}e^{3A}\tau' & .
\end{array}\right.
\label{floweq}
\ee
By performing a Legendre transformation of the form
\be
H_{\textrm{eff}} \,=\,A'\pi_A+s'\pi_s+\sigma'\pi_{\sigma}+t'\pi_{t}+\tau'\pi_{\tau} \,-\, L_{\textrm{eff}}(A,s,\sigma,t,\tau,A',s',\sigma',t',\tau') \ ,
\ee
one obtains the following 1D effective Hamiltonian
\be
\label{Heff_FlatDW}
H_{\textrm{eff}} \,=\, e^{-3 A} \left(\frac{\pi_A^2}{12}-\frac{\sigma^2}{4} (\pi_{s}^2+\pi_{\sigma}^2) -\frac{\tau^2}{3} (\pi_{t}^2+\pi_{\tau}^2) \right)\,+\,e^{3 A}\,V(s,\sigma ,t,\tau)\ .
\ee
According to the HJ formalism, the second order dynamics associated with the Lagrangian \eqref{Leff_FlatDW} turns out be equivalent to the following first order flow (where $\{\phi^{I}\} \equiv \{s,\sigma,t,\tau\}$)
\be\left\{
\begin{array}{lclc}
\pi_{A} & = & \dfrac{\partial F}{\partial A} \, \equiv \,  F_A & , \\[3mm]
\pi_{I} & = & \dfrac{\partial F}{\partial \phi^{I}} \, \equiv \,  F_{I}& , 
\end{array}\right. \label{HJ_flow_eqns}
\ee
when supplemented with the HJ constraint $H_{\textrm{eff}}|_{\pi_I\rightarrow F_I}\,\overset{!}{=}\,0$, which fixes the form of the generating function $F(A,s,\sigma ,t,\tau)$. In this case, the HJ constraint takes the following explicit form
\be\label{HJ_Constraint}
 e^{-3 A} \left(\frac{F_A^2}{12}-\frac{\sigma^2}{4} (F_{s}^2+F_{\sigma}^2) -\frac{\tau^2}{3} (F_{t}^2+F_{\tau}^2) \right)\,+\,\underbrace{e^{3 A}\,V(s,\sigma, t,\tau)}_{V_{\textrm{eff}}(A,s,\sigma, t,\tau)}\,\overset{!}{=}\,0 \ .
\ee
Since the effective potential has a \emph{factorized} dependence of the warp factor $A$, the corresponding generating functional may be found by casting a separable \emph{Ansatz} $F(A,s,\sigma, t,\tau)\,=\, 2 e^{3A}\,f(s,\sigma, t,\tau)$, where the function $f$ of the scalar fields satisfies the following non-linear PDE
\be
\label{PDE_f}
-3f^2 \,+\, 2K^{ij}\partial_i f\partial_j f\,\overset{!}{=}\, V(s,\sigma ,t,\tau) \ ,
\ee
where $K^{ij}\,\equiv\, \textrm{diag}\left(\frac{\sigma^2}{2},\,\frac{\sigma^2}{2},\,\frac{2\tau^2}{3},\,\frac{2\tau^2}{3}\right)$ is nothing but the inverse kinetic metric.
This identity manifestly shows the interpretation of these first order flows as a realization of \emph{fake supergravity} \cite{Freedman:2003ax}.

Indeed, one can show that 
\begin{multline}
f_{\textrm{SUSY}} \, \equiv \, e^{K/2}\,\left| W\right| \, = \, \frac{1}{4 \sqrt{2}\sigma^2 \tau ^{3/2}} \big( (1 + 6 s^2 t + t^3 - 6 t \sigma ^2 - 12 s \sigma \tau - 
   3 t \tau^2 )^2 \\
   +( -12 s t \sigma - 
   3 (2 s^2 + t^2 - 
      2 \sigma^2) \tau + \tau ^3 )^2
\big) ^{1/2}~.
\end{multline}
is a global solution to the PDE \eqref{PDE_f}, corresponding to the actual superpotential of the theory. Thanks to the existence of the superpotential, one may construct a globally bounding function as $-3f_{\textrm{SUSY}}^2$, which dynamically protects the SUSY vacuum from non-perturbative decay by virtue of the positive energy theorem for supersymmetry \cite{Witten:1982df} (see also \cite{Townsend:1984iu,Nester:1993fn}).

\section{Positive energy theorems for non-SUSY vacua}
\label{Sec5}

Now, because the PDE \eqref{PDE_f} characterizing the function $f$ is non-linear, $f_{\textrm{SUSY}}$ may not be the only global solution. In \cite{Boucher:1984yx} it was studied how solutions of the aforemoentioned PDE can imply dynamical protection from non-perturbative decay. The claim can be summarized into the following theorem \cite{Breitenlohner:1982bm}.

\textbf{Theorem:} If a scalar potential $V(\phi^i)$ in a given EFT can be written as
\be
V\,=\,-3f^2 \,+\, 2K^{ij}\partial_i f\partial_j f \ ,
\ee
for a suitable globally defined function $f$ of the scalars such that:
\begin{itemize}
\item $(i)$ $\partial_{i}f|_{\phi_0}\,=\,0\,$,
\item $(ii)$  $V(\phi) \, \geq \, -3f(\phi)^2$, $\qquad \forall \,\phi \in \mathcal{M}_{\textrm{scalar}}$ ,
\end{itemize}
 then other points in the scalar manifold $\mathcal{M}_{\textrm{scalar}}$ have higher energy than $\phi_0$ itself, whence $\phi_0$ is stable against non-perturbative decay within the given EFT.  

Note that condition $(ii)$ requires complete knowledge concerning the global properties of $f$ on the scalar manifold, and hence it is in general very difficult to check. What is usually done in practice is to combine local patches with the asymptotic form of the solutions which are valid at the boundary of $\mathcal{M}_{\textrm{scalar}}$. This allows to exclude unexpected behaviors in intermediate regions.
As remarked earlier, due to the non-linearity of the defining PDE, multiple global $f$'s might exist. Moreover, since any critical point would represent a \emph{singular} initial condition, multiple solutions are possible even at a local level.
By following the analysis in \cite{Danielsson:2016rmq}, one may construct all of the local solutions of \eqref{PDE_f} by perturbatively expanding around each of the critical points of the scalar potential. This way, we may look for explicit solutions of the form\footnote{In this general formalism we collectively denote the scalar fields by $\phi\,\equiv\,(s,\sigma,t,\tau)$. As a consequence, the $k$-th derivative of a scalar function $f$ is represented by a rank $k$ tensor field.}
\be
f = f^{(0)} +\frac{1}{2}(\phi^{i}-\phi^{i}_0)(\phi^{j}-\phi^{j}_0)f^{(2)}_{ij}+\frac{1}{3!}(\phi^{i}-\phi^{i}_0)(\phi^{j}-\phi^{j}_0)(\phi^{k}-\phi^{k}_0)f^{(3)}_{ijk}+...~,
\ee
where $f^{(k)}_{i_1....i_k}=\tfrac{\partial ^k f}{\partial \phi^{i_1}...\partial \phi^{i_k}}|_{\phi_0}$ is the generalised $k$-th derivative of $f$ taken at $\phi_0$. After fixing the 0th and 1st order coefficients to be 
\be
f^{(0)} \ = \ \sqrt{-\frac{V(\phi_0)}{3}} \ , \qquad  \textrm{and} \qquad f^{(1)} \ = \ 0 \ ,
\ee
the second order coefficients satisfy second degree algebraic equations admitting $2^4=16$ different branches of solutions. After fixing this discrete choice at 2nd order, the whole tower of higher order coefficients will be completely and uniquely determined by just solving degree 1 algebraic equations.

\begin{figure}[h!]
	\centering
 \includegraphics[width=8.5cm,trim = 0cm 0cm 0cm 1.5cm,
  clip]{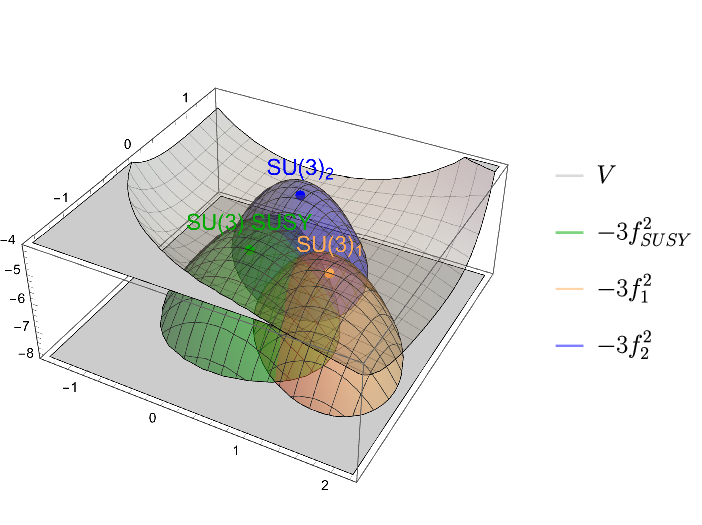}
 \caption{\it The profile of the scalar potential of $\mathrm{SU}(3)$ invariant $\mathrm{ISO}(7)$ gauged supergravity (grey surface), against three globally bounding functions $-3f^2$, obtained by solving the PDE \eqref{PDE_f} through a perturbative expansion around the three critical points with $\mathrm{SU}(3)$ residual symmetry.}
\label{fig:POsitive_Energy_Proof}
\end{figure}

As required by the above argument, the particular choice of local $f$ having a global extremum at a given $\phi_0$ is relevant for discussing positive energy theorems and non-perturbative stability of the related vacuum. Amongst the 16 possible branches at each vacua, they correspond to the the choice of Hessian with the highest eigenvalues. As an example, we plot in figure \ref{fig:POsitive_Energy_Proof} the three bounding functions ensuring the global stability of the three SU(3) points within the 2D-slice containing these three points. The fake superpotentials plotted there are global functions in the sense that was anticipated above, \emph{i.e.} the local behavior in the vicinity of each critical point smoothly glues to the desired asymptotiic behavior at infinity.

The first order flows defined by each local $f$ may end in three different ways: at a singularity (only if $f$ may not be globally extended!), at infinity (as for the ones in figure \ref{fig:POsitive_Energy_Proof}), or at a BF-stable AdS extremum. DWs are precisely flows that have both end points of this third type.
In order to find DWs, we need to look for global $f$'s that have a very flat profile in field space.
For this reason, the choices of Hessian with the lowest eigenvalues turn out to define flat DW solutions interpolating between different vacua. In terms of $\Delta$ modes, the ones with the lowest values in table \ref{Table:Mass_spectra} parametrise DWs. They are colored according to the colors used later in figure \ref{DWmap}.

The perturbative analysis has been carried out up to order $20$, in order to correctly integrate the flow equations \eqref{floweq}.  

\section{Interpolating flat DW solutions}\label{Sec:DW_Plots}

\begin{figure}[h]
\centering
\begin{tikzpicture}[scale=0.9,
    node distance=2cm,
    group/.style={draw=gray, align=center,rounded corners=5pt},
    arrow/.style={-, thick}
]

% Définition des nœuds principaux (respectant les hauteurs relatives de l'image)

\node[group] (SO7) at (0, 4) {$SO(7)_+$\\$\mathcal{N}=0$};
\node[group] (SO6) at (-3.7, 0.8) {$SO(6)_+$\\$\mathcal{N}=0$};
\node[group] (G2) at (1, 2.2) {$G_2$\\$\mathcal{N}=1$};
\node[group] (SU3U1) at (-1.2, 1.2) {$SU(3) \times U(1)$\\$\mathcal{N}=2$};
\node[group] (SU3) at (0.6, -1.3) {$SU(3)$\\$\mathcal{N}=1$};
\node[group] (G2star) at (4.7, 0.5) {$G_2^*$\\$\mathcal{N}=0$};
\node[group] (SU3_2) at (2.1, 0.3) {$SU(3)_2$\\$\mathcal{N}=0$};
\node[group] (SU3_1) at (3.4, -1) {$SU(3)_1$\\$\mathcal{N}=0$};

% Flèches de connexion
\draw[arrow,red] (SO7) -- (SO6);
\draw[arrow] (SO7) -- (G2star);
\draw[arrow] (SO7) -- (G2);
\draw[arrow,red] (SO6) -- (SU3U1);
\draw[arrow,green!60!black] (SO6) -- (SU3);
\draw[arrow] (G2) -- (SU3U1);
\draw[arrow] (G2) -- (SU3);
\draw[arrow,blue] (G2star) -- (SU3_2);
\draw[arrow,orange!92!black] (G2star) -- (SU3_1);
\draw[arrow,green!60!black] (SU3_2) -- (SU3);
\draw[arrow,green!60!black] (SU3) -- (SU3_1);
\draw[arrow,green!60!black] (SU3) -- (G2star);
%\draw[arrow,green!60!black] (0.8, -0.8) -- (1.1, 1.7);

\end{tikzpicture}
\caption{\it Domain wall net of the SU(3) invariant sector. Each pair of boxes connected by a line represents a pair of AdS critical points connected by a DW. The colors match the colors used in Table.\ref{Table:Mass_spectra} to label the modes triggered for the corresponding DW.  Equivalently, a color labels the fake superpotential $f$ used to integrate the flow equations.  }
\label{DWmap}
\end{figure}
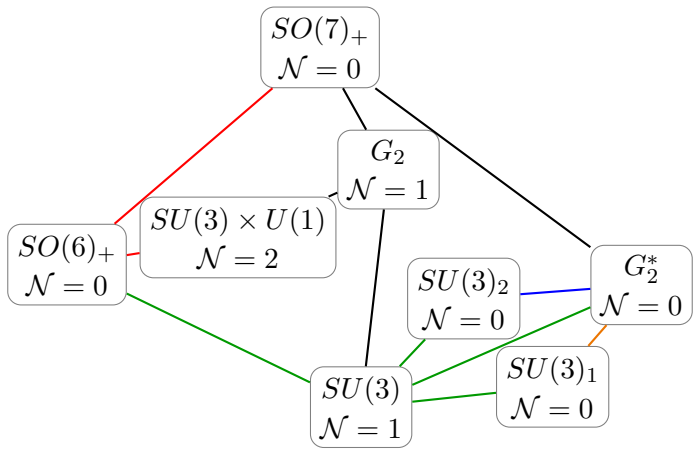

In the SU(3) invariant sector, the known DWs linking SUSY vacua have been computed again and new ones were found between non-SUSY points by using the perturbative method. First, before providing the details of the DWs found, we will take a look at the concluding map of DWs figure \ref{DWmap}. All the black lines represent DWs that were already known and the coloured ones are the new ones, which include SUSY to non-SUSY but also non-SUSY to non-SUSY DWs. The different colors correspond to different fake superpotentials $f$, the most predominant one being the branch coming from the SU(3) $\&~\mathcal{N}=1$ vacuum. The vacua are also placed vertically according to their potential value (without scale) as in Table~\ref{Table:Critical_Points} to show the direction of the flows \cite{alvarez2000holographyctheorem}. All the vacua represented have $t>0$ because it happens they grouped together. The rest of the map with the points having $t<0$ is the symmetric with respect to the axis passing by SO(7) and SO(6) (SU(3)$\times$U(1) should also be on this line).

Surprisingly, the BF unstable SO(7) vacuum has 3 DWs connecting to it despite its tachyon. This is because the 2 imaginary modes corresponding to the mass below the BF bound are not excited along these DWs meaning in this sector this tachyon is frozen, but if we zoom out to a larger sector such as the whole $\mathcal{N}=8$ theory, then these domain walls would appear unstable for any perturbation along the direction of the tachyon. What is most puzzling in this map is the fact some points are not linked when one could expect them to be. For example, no DWs can be found between $G_2~ \&~\mathcal{N}=1$ and $G_2~ \&~\mathcal{N}=0$, nor between SU(3) $\&~ \mathcal{N}=1$ and SU(3)$\times$U(1) $\&~ \mathcal{N}=2$ despite the superpotential $f_{\textrm{SUSY}}$ linking them. We shall see that this can be explained by the impossibility of finding a monotonic (fake) superpotential interpolating between these points.

\begin{figure}[h!]
\centering
\begin{subfigure}[b]{0.49\textwidth}
        \includegraphics[width=\textwidth]{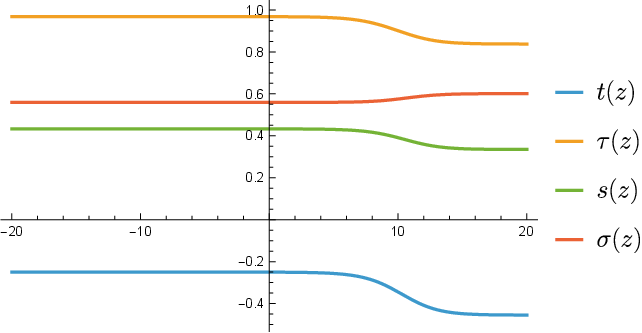}
        \subcaption{SU(3) $\&~\mathcal{N}=1$ $\rightarrow$ SU(3)$_1$}
    
    \end{subfigure}
    \hfill
\begin{subfigure}[b]{0.49\textwidth}
        \includegraphics[width=\textwidth]{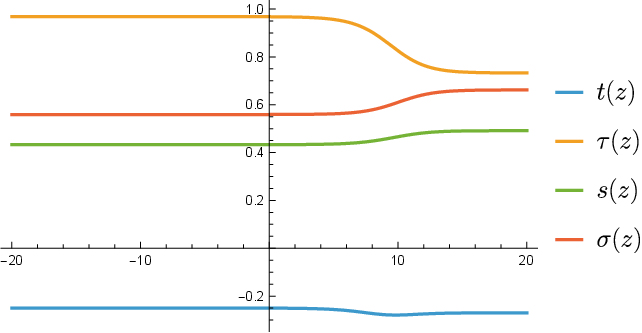}
        \subcaption{SU(3) $\&~\mathcal{N}=1$ $\rightarrow$ SU(3)$_2$}
     \end{subfigure}
   \vskip\baselineskip   
   \begin{subfigure}[b]{0.49\textwidth}
        \includegraphics[width=\textwidth]{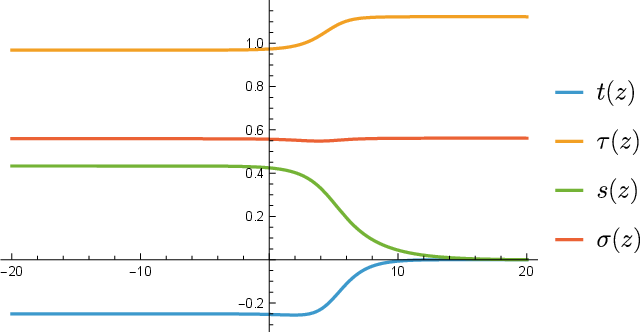}
        \subcaption{SU(3) $\&~\mathcal{N}=1$ $\rightarrow$ SO(6)$_+$}
      
    \end{subfigure} 
      \hfill
\begin{subfigure}[b]{0.49\textwidth}
        \includegraphics[width=\textwidth]{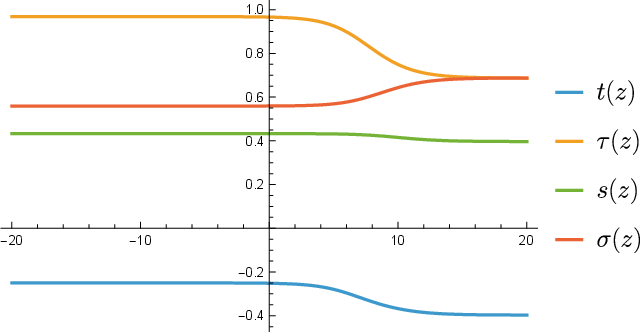}
        \subcaption{SU(3) $\&~\mathcal{N}=1$ $\rightarrow$ $G_2^*~\&~\mathcal{N}=0$}
       
    \end{subfigure}
       \vskip\baselineskip  
    \begin{subfigure}[b]{0.49\textwidth}
        \includegraphics[width=\textwidth]{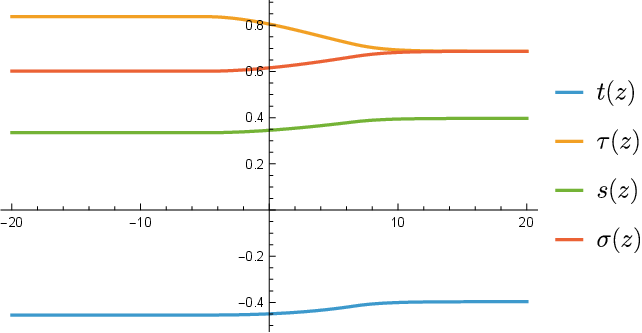}
        \caption{SU(3)$_1$ $\&~\mathcal{N}=0$ $\rightarrow$ $G_2^*~\&~\mathcal{N}=0$}
    \end{subfigure}
    \hfill
\begin{subfigure}[b]{0.49\textwidth}
        \includegraphics[width=\textwidth]{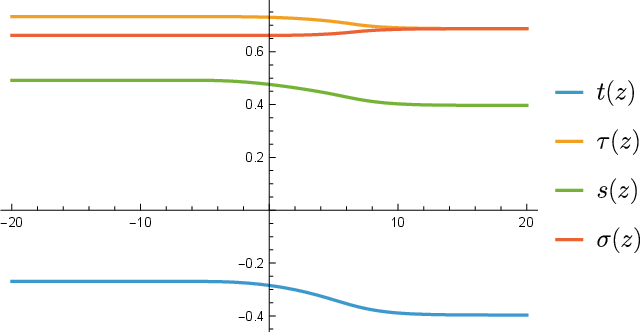}
        \caption{SU(3)$_2$ $\&~\mathcal{N}=0$ $\rightarrow$ $G_2^*~\&~\mathcal{N}=0$}   
    \end{subfigure}
     \vskip\baselineskip
    \begin{subfigure}[b]{0.49\textwidth}
        \includegraphics[width=\textwidth]{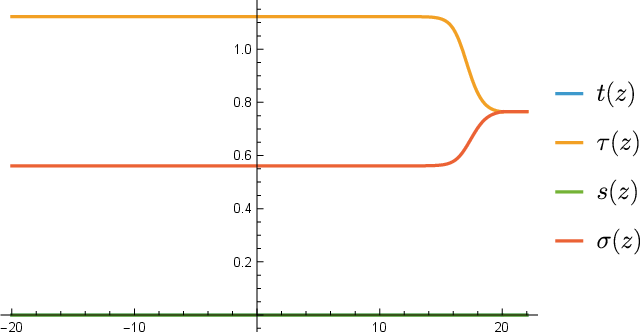}
        \caption{SO(6)$_+$ $\rightarrow$ SO(7)$_+$}
        
    \end{subfigure}
   \hfill
    \begin{subfigure}[b]{0.49\textwidth}
        \includegraphics[width=\textwidth]{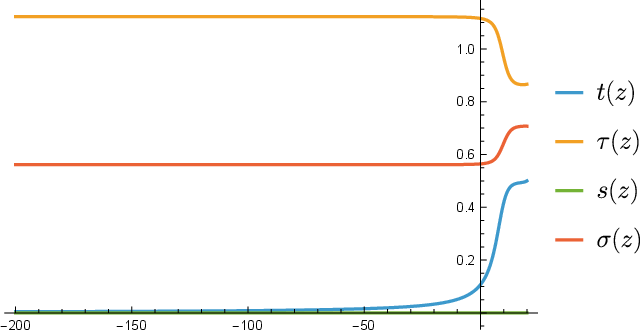}
        \caption{SO(6)$_+$ $\rightarrow$ SU(3)$\times$ U(1)}
        
    \end{subfigure}
	\caption{\it Profiles of the fields $(t,\tau,s,\sigma)$ for all eight DWs colored in Figure~\ref{DWmap}. Notice some DWs originate from the same fake superpotential but correspond to different boundary conditions according to the different AdS vacua they interpolate between.}
	\label{profiles}
\end{figure}

Let's explore in more detail these DWs. Fig.\ref{profiles}(a to d) shows the profiles of the scalar fields along the DWs originating from SU(3) $\&~\mathcal{N}=1$ (in green in Fig.\ref{DWmap}). All these solutions to the flow equations come from the same (green) fake superpotential corresponding to the excited modes $\{1-\sqrt{6}~(\times 2),2-\sqrt{6}~(\times 2) \}$. Similar profiles are obtained (Fig.\ref{profiles}(e,f)) for the DWs ending in $G_2^*~\&~\mathcal{N}=0$ (in blue and yellow in Fig.\ref{DWmap} and in Table\ref{Table:Mass_spectra} for the modes) and for the DWs starting in SO(6) (Fig.\ref{profiles}(g,h) and in red in Fig.\ref{DWmap} and in Table\ref{Table:Mass_spectra}).

\begin{figure}[h!]
\centering
\begin{subfigure}[t]{0.55\textwidth}
      \includegraphics[width=\textwidth]{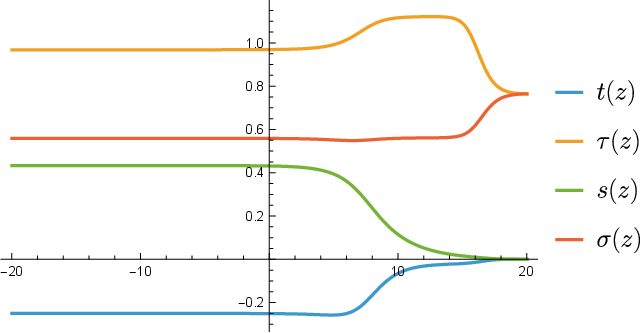}
        \caption{}        
    \end{subfigure}
\hfill
  \begin{subfigure}[t]{0.38\textwidth}
      \includegraphics[width=\textwidth]{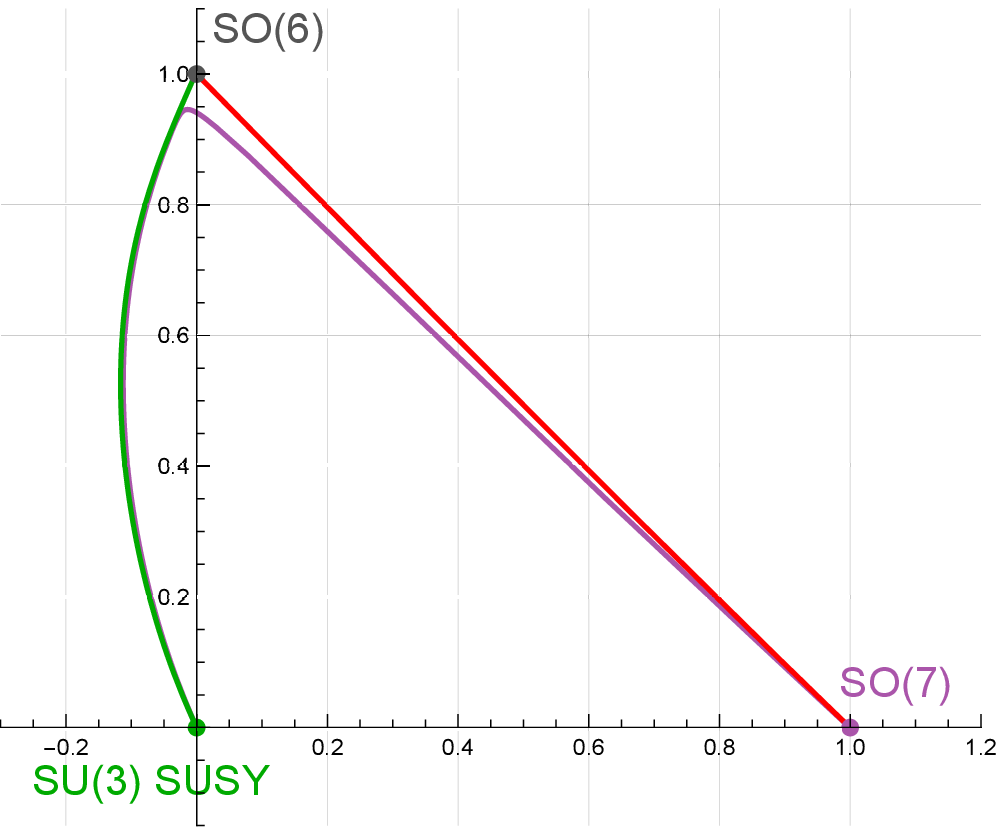}
        \caption{}        
    \end{subfigure}

\caption{\textit{Profiles of the fields for the DW SU(3) $\&~\mathcal{N}=1$ $\rightarrow$ SO(7)$_+$ (a) and trajectories of this DW (in purple although it uses the green fake superpotential) and of the SU(3) $\&~\mathcal{N}=1$ $\rightarrow$ SO(7)$_+$ (green) and SO(6)$_+$ $\rightarrow$ SO(7)$_+$ (red) in the 2D-slice of the scalar manifold that contains these vacua.}}
\label{doublewall}
\end{figure}

An extra interesting wall was found using the green fake superpotential. It starts from SU(3) $\&~\mathcal{N}=1$, nearly passes through SO(6)$_+$ and ends in SO(7)$_+$ (Fig~\ref{doublewall}). We can see that the profiles of the fields Fig~\ref{doublewall}(a) look like the exact combination of the graphs Fig~\ref{profiles}(c) and Fig~\ref{profiles}(g), yet the fields do not exactly pass through the SO(6) invariant point, as it can be seen on the plot Fig~\ref{doublewall}(b). It seems that the greater the precision of the numerical calculations is, the closer the path will pass near the SO(6) point, meaning both paths are in fact the same.  

\begin{figure}[h!]
\centering
\begin{subfigure}[b]{0.46\textwidth}
      \includegraphics[width=\textwidth]{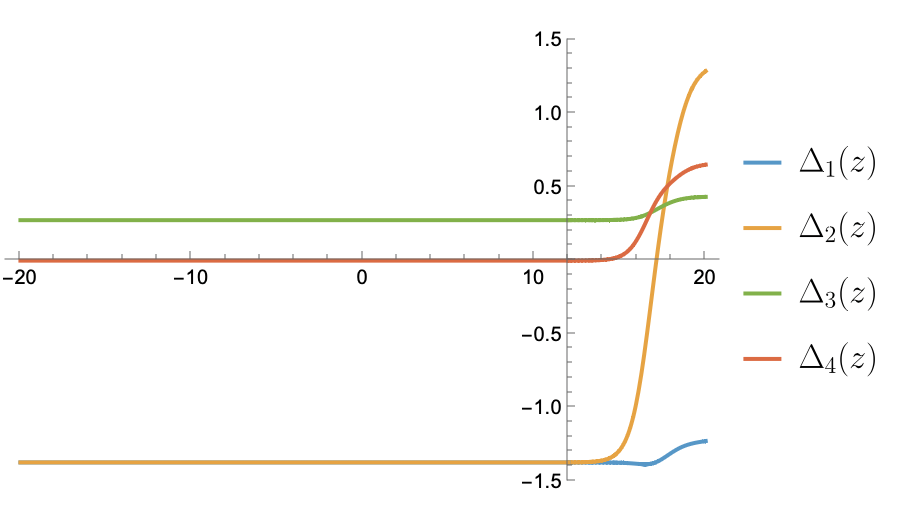}
\caption{ SO(6)$_+$ $\&~\mathcal{N}=0$ $\rightarrow$ SO(7)$_+$ }        
    \end{subfigure}
    \hfill
\begin{subfigure}[b]{0.46\textwidth}
        \includegraphics[width=\textwidth]{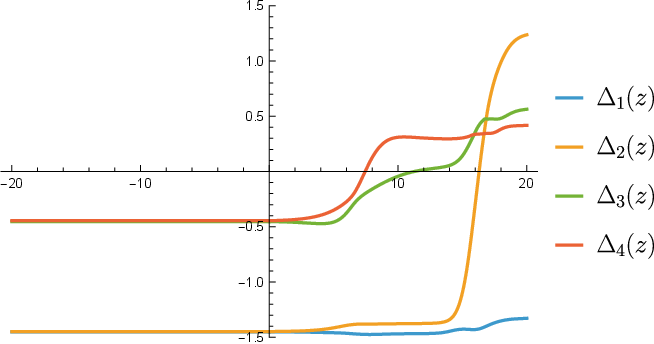}
        \caption{SU(3) $\&~\mathcal{N}=1$ $\rightarrow$ SO(7)$_+$}
            \end{subfigure}
\begin{subfigure}[b]{0.46\textwidth}
        \includegraphics[width=\textwidth]{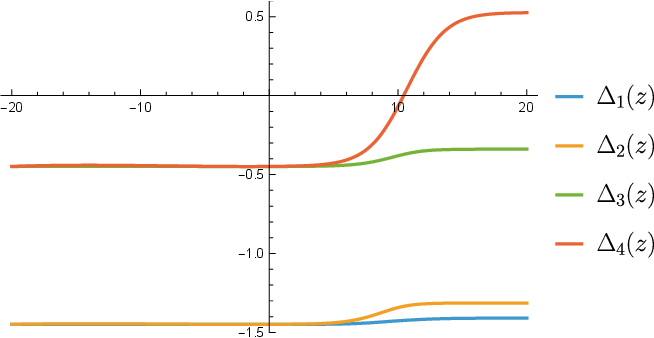}
        \caption{SU(3) $\&~\mathcal{N}=1$ $\rightarrow$ SU(3)$_1$($\mathcal{N}=0$)}
            \end{subfigure}
\caption{\it Evolution of the modes along the DWs ending in SO(7)$_+$ ((a) and (b)) and the DW: SU(3) $\&~\mathcal{N}=1$ $\rightarrow$ SU(3)$_1$ ($\mathcal{N}=0$) }
\label{deltas}
\end{figure}

In order to better understand why it is allowed to have DWs ending in the SO(7) vacuum, we can keep track of the modes triggering the flow along the DW by computing 
\begin{equation}
\Delta (z) = \text{Eigenvalues}\left[\left.2\sqrt{\tfrac{-3}{V}}f^{(2)}K^{-1}\right|_{\phi (z)}\right]\ . \nonumber
\end{equation} 
Fig~\ref{deltas} shows the evolution of the modes for both walls ending in SO(7). Again, we can see the double wall configuration by observing that both graphs have approximatively the same ending in the UV. DW (a) starts at the red modes (see Table~\ref{Table:Mass_spectra}) and DW (b) at the green ones, yet when arriving in SO(7) the modes do not correspond to any of the SO(7) modes in Table~\ref{Table:Mass_spectra} meaning they are one-way flows (one could not find these flows by perturbating the SO(7) point). The evolution of the modes for the DW SU(3) $\&~\mathcal{N}=1$ $\rightarrow$ SU(3)$_1$($\mathcal{N}=0$) is also given as a simpler example. For this case, the modes nicely interpolates between the values in green and the values in yellow in Table \ref{Table:Mass_spectra}.

\begin{figure}[h!]
\centering
       \includegraphics[width=9.5cm]{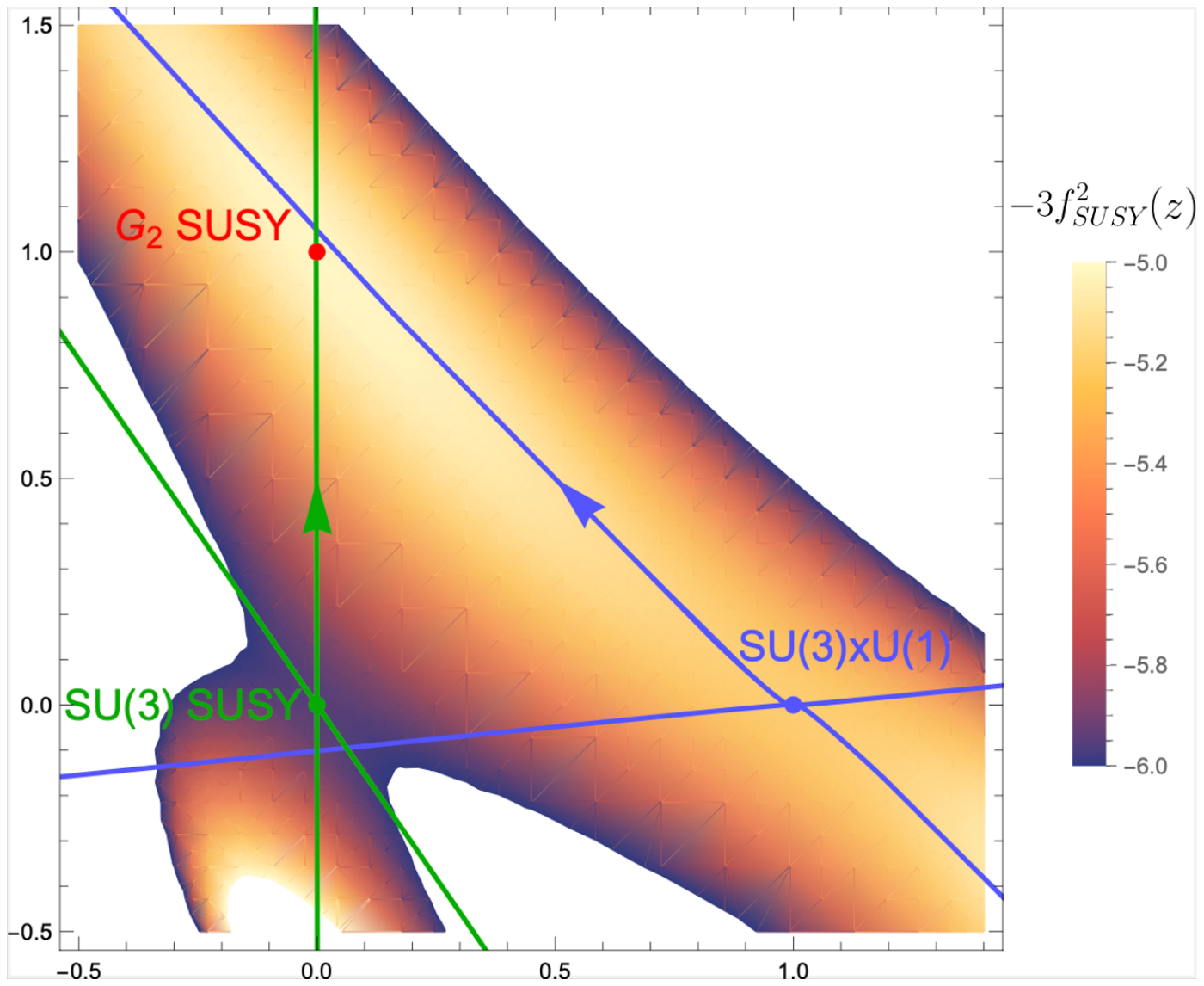}

\caption{\it Directions of the small perturbations of the IR points, plotted on the 2D-slice containing all three SUSY vacua SU(3), SU(3)$\times$U(1) and $G_2$.}
\label{WSUSY}
\end{figure}

No DW can be found between SU(3) $\&~\mathcal{N}=1$ and SU(3)$\times$U(1) $\&~\mathcal{N}=2$. The flattest fake superpotential, corresponding to the choice of Hessian with the most negative eigenvalues have been tested, but none of them interpolates between the two vacua (at least at a maximum Taylor expansion order of 20), with the exception of one, the superpotential $f_{\textrm{SUSY}}$ as expected. Unfortunately, the usual technique of integration of the fields that uses the fact that the gradient of $f$ is 0 in the IR and close to 0 in the UV, does not work here because the superpotential $f_{\textrm{SUSY}}$ being analytical, its gradient is perfectly 0 in both vacua. Expanding $f_{\textrm{SUSY}}$ in Taylor series around the IR point to approximate the gradient to nearly 0 in the UV has been tried but in vain since, being constrained by the time of the calculations, the order of the expansion was not pushed high enough to correctly achieve what was wanted. Wondering if the problem is a matter of monotonicity of the superpotential, we can plot $-3f_{\textrm{SUSY}}^2$ in the 2D-slices containing  the three SUSY points (Fig~\ref{WSUSY}).

It appears that the SU(3) point is a minimum, SU(3)$\times$U(1) is a saddle and $G_2$ is a maximum, so a monotonic path between the first two should be possible, just as it is for the walls connecting each of them to the $G_2$ point. Yet by solving the linearised flow equations \eqref{lineareq}, we can see in which direction the perturbations go (blue and green lines), and observe that a trajectory starting in the SU(3) point goes monotonically towards $G_2$ and not towards SU(3)$\times$U(1). Assuming the first direction gives the DW towards $G_2$, and as the second one necessarily has to go down, no trajectory can link the SU(3) and the SU(3)$\times$U(1) vacua. Of course, we cannot exclude the possibility that these vacua be linked through a trajectory that excites scalars outside of the SU$(3)$ invariant sector, which remains a case to be explored.

\section{Holographic Calculations \& Concluding Remarks}
\label{Sec:Holography}
\begin{figure}[h]
\centering
\begin{tikzpicture}[scale=1, every node/.style={scale=1}, 
    arrow/.style={-{Latex[length=2mm]}, thick},
    baseline=(current bounding box.north)]

  \draw[arrow] (-3,1.5) -- (3,1.5) ;
  \node at (0,1.8) {RG flow};  
  \node at (-3.1,1.2) {IR};
  \node at (3.1,1.2) {UV};
  \node at (3.1,1.8) {(energy $\mu$)};
  \node at (-3.1,0.8) {$\beta = 0$};
  \node at (3.1,0.8) {$\beta = 0$};
  \node at (0,1.1) {QFT$_d$};
  \node at (0,0.7) {$\beta \neq 0$};

  \draw[arrow] (-3,-1) -- (3,-1) node[midway, above] {DW$_{d+1}$};
  \node at (-3.1,-1.4) {AdS$^{\text{IR}}_{d+1}$};
  \node at (3.1,-1.4) {AdS$^{\text{UV}}_{d+1}$};
  \node at (3.1,-0.7) {(rad. coord. $z$)};
  \draw [->, thick,dashed] (-3.6,1.2) to[out=200, in=160] (-3.6,-0.9);
  \node at (-4.5,0.1) [rotate=90] {Holography};

\end{tikzpicture}
\caption{\textit{The DW/QFT correspondence summarized.}}
\label{holography}
\end{figure}
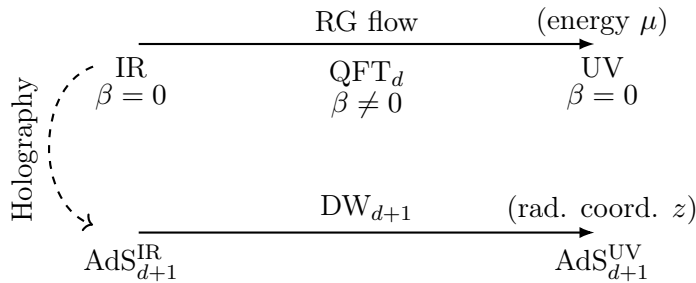

In this paper we discussed the existence of a net of DW solutions interpolating between different pairs of AdS vacua within the $\mathrm{SU}(3)$ invariant sector of $\mathrm{ISO}(7)$ gauged supergravity, which turn out to be solutions of massive IIA supergravity, by virtue of the uplift formulae in appendix. While the supersymmetric DWs were already found in \cite{Guarino:2016ynd}, we presented a few non-supersymmetric DWs that are novel, at least to our knowledge. A summary of the web of flows connecting the different vacua is depicted in Fig.\ref{DWmap}. 
In the light of the DW/QFT correspondence, each of these would be dual to an RG flow within a 3d quantum field theory, with different conformal fixed points in the UV and in the IR. This situation is schematically represented in Fig.\ref{holography}. 

In 3d, quantum field theories are believed to obey the F-theorem \cite{Pufu:2016zxm}, stating that $F_{\mathrm{UV}}\,>F_{\mathrm{IR}}$, $F$ denoting the free energy. This statement agrees with the common intuition that the effective number of degrees of freedom which are excited at a given energy scale monotonically increases with the energy scale itself.
Within the holographic setting discussed in our paper, the holographic free energy at each conformal fixed point is related to the AdS$_4$ radius of the dual vacuum expressed in Planck units, through
\begin{equation}
F_{\mathrm{hol}} \ \propto \ \frac{L^{2}}{\kappa_4^2} \ \propto \ \left.\frac{1}{f^2\kappa_4^2}\right|_{\textrm{flow}} \ ,
\end{equation}
where $f$ denotes the fake superpotential defining the corresponding interpolating DW solution. It is worth mentioning that this identification is consistent with a holographic F-theorem, in that $f$ is strictly monotic on every single flow. Following the same logic, the plots shown in \ref{deltas}, represent a sketch of the anomalous dimension picked up by each dynamical operator along the RG flow. Finally, the $\beta$ function associated with the deformation responsible for triggering the flow, is related to the radial derivative of the fake superpotential evaluated along the flow.
While similar plots may be produced for all the flows found in this paper, we only show the behavior of the holographic free energy and the corresponding holographic beta function for one particular case, just for the sake of brevity.

\begin{figure}[h]
\centering
      \includegraphics[width=7.5cm]{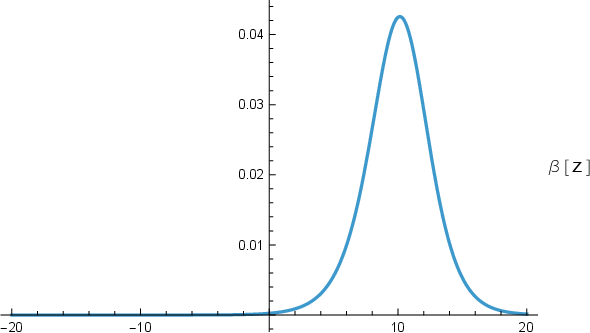}
      \includegraphics[width=7.5cm]{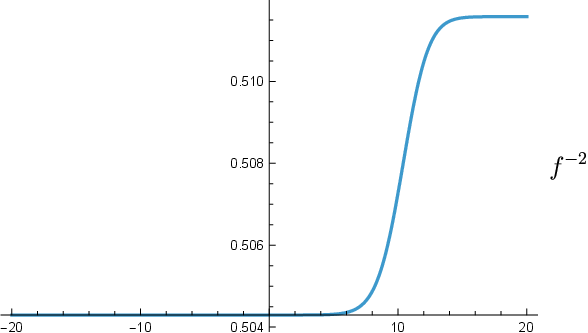}
        \caption{\textit{Holographic $\beta$-function and free energy for the flow SU(3)$\&~\mathcal{N}=1 \rightarrow$ SU(3)$_1$ ($\mathcal{N}=0$)}}   
        
\end{figure}

One final remark is due. Our present discussion concerning the possible relevance of our results for holography is motivated by the stability analysis done in \cite{Guarino:2020flh,Guarino:2020jwv} for the non-supersymmetric AdS vacua that we considered. We should stress that these findings are still partial and do not completely exclude the possible existence of an instability channel for these vacua, in agreement with the predictions of the AdS swampland conjecture. However, there could be three different scenarios that one might envision here. 
\begin{itemize}
\item A non-perturbative decay channel is indeed there for all non-supersymmetric AdS vacua that we considered and this automatically destroys the existence of any holographic dictionary.

\item A non-perturbative decay channel is indeed there for all non-supersymmetric AdS vacua that we considered, but one might still understand how to think of a holographic correspondence which is only valid below a certain cut-off scale fixed by the aforementioned instability.

\item No instabilities affect these vacua and they are strong candidates to become the first examples of non-supersymmetric holography within a controlled setup.
\end{itemize}

As of now, we are unable to actually tell which of the above options corresponds to the truth, but we definitely plan to keep investigating on this extremely challenging issue. One crucial test that might easily rule out the third option could be to take the data of the putative non-SUSY AdS vacua and use them to define a spectrum of low-lying operators of a dual CFT with a given set of conformal dimensions. With this input, one might search for solutions to the crossing symmetry equations by resorting to numerical bootstrap techniques. We hope to clarify this in the future.

\section*{Acknowledgements}

We would like to thank Valentina Bevilacqua, Adolfo Guarino and Giuseppe Sudano for some valuable comments on a draft version of this manuscript. 
NB also thanks the Physics Department of Rome Tor Vergata University for their warm hospitality while this project was completed.
The work of NB is supported by ENS (Ecole Normale Supérieure) de Lyon.

\appendix

%\section{Massive type IIA supergravity}

\section{Consistent truncations of massive IIA supergravity on $S^6$}\label{AppA}
In this appendix we include some details concerning the 10D origin of $\mathrm{SU}(3)$ invariant $\mathrm{ISO}(7)$ gauged supergravity in four dimensions.
Massive type IIA supergravity admits a consistent truncation on the six-sphere, for which the full non-linear KK \emph{Ansatz} was determined in \cite{Guarino:2015vca}. Its field content is given by the 4D metric $g_{\mu\nu}$; the complex scalar fields $S\,=\, s \, + \, i \sigma$ and $T\,=\, t \, + \, i \tau$; 2 electric vectors $A^0$ and $A^1$, and their magnetic duals $\widetilde{A}_0$ and $\widetilde{A}_1$; 3 two-forms $B^0$, $B_1$ and $B_2$, and 2 three-forms $C^0$ and $C_1$. The four real scalars are the remaining scalars after gauge fixing the six scalars parametrising the submanifold \eqref{scalarmanifold}. One of the two gauge-fixed scalar, the Stückelberg scalar $a$, appears in the expressions even though it is completely fixed by the parameter $B^0$ \cite{Guarino:2015qaa}.    

The full reduction \emph{Ansatz} for the 10D fields in the \emph{string frame} reads
\begin{equation}
\label{SU3ISO7Truncation}
\begin{split}
d s_{10}^2&=\frac{\widetilde{\Delta}_2^{1/2}}{\sigma^2\tau}ds^2_4+g^{-2}\frac{\widetilde{\Delta}_2}{\tau^2+t^2}\left[d\alpha^2+\sin^2\alpha\left(\frac{\tau^2}{\widetilde{\Delta}_2}\boldsymbol{\widetilde{\eta}}^2+ \frac{\tau^2+t^2}{\widetilde{\Delta}_1}ds^2_{\mathbb{CP}^2}\right)\right] \ ,\\
e^{\Phi}&= \frac{\widetilde{\Delta}_2^{3/4}}{\sigma(\tau^2+t^2)^{1/2}}\widetilde{\Delta}_1^{-1}\ , \\
\hat{B}_{(2)}&= -\cos\alpha\, B^0 +g^{-1}\sin\alpha\,\widetilde{A}_0\wedge d\alpha -g^{-2} \big{[}t(\tau^2+t^2)^{-1}\sin\alpha d\alpha \wedge \boldsymbol{\widetilde{\eta}} +t\widetilde{\Delta}_1^{-1}\sin^2\alpha\cos\alpha \,\boldsymbol{J} \\
&+s \widetilde{\Delta}_1^{-1} \sin^3\alpha\, \Re \,\boldsymbol{\Omega} \big{]} \ , \\
\hat{A}_{(1)}&= -\cos\alpha \,A^0-g^{-1}a\sin\alpha d\alpha + g^{-1}(\sigma^2t^2-\tau^2s^2)\widetilde{\Delta}_2^{-1}\sin^2\alpha\cos\alpha \,\boldsymbol{\widetilde{\eta}} \ , \\
\hat{A}_{(3)}&= \cos ^2 \alpha \, C^0+\sin ^2 \alpha \,C^1 -g^{-1} \sin \alpha \cos \alpha\left(B_1+\frac{3}{2} A^0 \wedge \widetilde{A}_0+\frac{1}{6} A^1 \wedge \widetilde{A}_1\right) \wedge d \alpha \\
&-\frac{1}{3} g^{-1} \sin ^2 \alpha \, B_2 \wedge \boldsymbol{\widetilde{\eta}} +\frac{1}{3} g^{-2} \widetilde{A}_1 \wedge\left[\sin ^2 \alpha \,\boldsymbol{J}+\sin \alpha \cos \alpha d \alpha \wedge \boldsymbol{\widetilde{\eta}}\right]-\cos \alpha \, A^0 \wedge \hat{B}_{(2)} \\
&+g^{-3}\widetilde{\Delta}_1^{-1}\big{[} a\big{(}t\sin^3\alpha\cos\alpha \,\boldsymbol{J}\wedge d\alpha +s\sin^2\alpha \,\Re\,\boldsymbol{\Omega}\wedge d\alpha \big{)} + t(\sigma^2+s^2)\sin^4\alpha \,\boldsymbol{J}\wedge \boldsymbol{\widetilde{\eta}} \\
&-s(\tau^2+t^2)\sin^3\alpha\cos\alpha \, \Re \,\boldsymbol{\Omega} \wedge \boldsymbol{\widetilde{\eta}}-s\sin^2\alpha \, \Im \,\boldsymbol{\Omega}\wedge d\alpha \big{]} \ ,
\end{split}
\end{equation}
where $ds_{4}^2$ denotes the 4D metric, $\alpha$ is an angle on $S^6$, $ds^2_{\mathbb{CP}^2}$ is the Fubini-Study metric on the $\mathbb{CP}^2$ base of $S^5$ within $S^6$, and $\boldsymbol{\widetilde{\eta}}=\boldsymbol{\eta}+gA^1$, with $\boldsymbol{J}$,  $\boldsymbol{\eta}$ and $\boldsymbol{\Omega}$ defining a Sasaki-Einstein structure of $S^5$ within $S^6$. The warp factors $\widetilde{\Delta}_1$ \& $\widetilde{\Delta}_2$ are defined by
\begin{equation}
\begin{array}{lcccclc}
\widetilde{\Delta}_1 =(\tau^2+t^2)\cos^2\alpha+(\sigma^2+s^2)\sin^2\alpha & , & & \textrm{and} & & \widetilde{\Delta}_2=\tau^2\cos^2\alpha+\sigma^2\sin^2\alpha & .
\end{array}\nonumber
\end{equation}

The consistency of the truncation automatically guarantees the equivalence between the 4D equations of motion obtained from the effective Lagrangian \eqref{Leff_FlatDW} of $\mathrm{ISO}(7)$ gauged supergravity and the 10D equations of motion of massive IIA supergravity.
In particular, by virtue of the uplift formulae given in \eqref{SU3ISO7Truncation}, the eight vacuum solution collected in table \ref{Table:Critical_Points} may be directly interpreted as $\mathrm{AdS}_4\times S^6$ solutions of massive IIA supergravity. 

 \bibliographystyle{utphys}
 \bibliography{BD25.bib}
\end{document}